\documentclass[twocolumn,letterpaper,10pt]{IEEEtran}
\usepackage{amsmath,epsfig}
\usepackage{bm}
\usepackage{latexsym}
\usepackage{amsmath}
\usepackage{amssymb}
\usepackage{graphicx}
\usepackage[algonl,boxed]{algorithm2e}

\begin{document}
%


\title{ Relay-Assisted Partial Packet Recovery with IDMA Method in CDMA Wireless Network\vspace{-3mm}}

\author{
Zhifeng Luo, Zhu Han$^*$, Albert Kai-sun Wong$^\dag$, and Shuisheng Qiu\\
\authorblockA{School of Electronic and Information Engineering,South China University of Technology,Guangzhou, China\\
$^*$  Electrical and Computer Engineering Department, University of Houston, Houston, TX, USA\\
$^\dag$   Department of Electronic and Computer Engineering, University of Science and Technology, Hong Kong, China
}\thanks{This work is supported by US NSF CNS-0953377, CNS-0905556 and CNS-0910461.}\vspace{-9mm}
}

\maketitle
\begin{abstract}
Automatic Repeat Request (ARQ) is an effective technique for
reliable transmission of packets in wireless networks. In ARQ,
however, only a few erroneous bits in a packet will cause the entire
packet to be discarded at the receiver. In this case, it's wasteful
to retransmit the correct bit in the received packet. The partial
packet recovery only retransmits the unreliable decoded bits in
order to increase the throughput of network. In addition, the
cooperative transmission based on Interleave-division
multiple-access (IDMA) can obtain diversity gains with multiple
relays with different locations for multiple sources simultaneously.
By exploring the diversity from the channel between relay and
destination, we propose a relay-assisted partial packet recovery in
CDMA wireless network to improve the performance of throughput. In
the proposed scheme, asynchronous IDMA iterative chip-by-chip
multiuser detection is utilized as a method of multiple partial
recovery, which can be a complementarity in a current CDMA network.
The confidence values' concept is applied to detect unreliable
decoded bits. According to the result of unreliable decoded bits'
position, we use a recursive algorithm based on cost evaluation to
decide a feedback strategy. Then the feedback request with minimum
cost can be obtained. The simulation results show that the
performance of throughput can be significantly improved with our
scheme, compared with traditional ARQ scheme. The upper bound with
our scheme is provided in our simulation. Moreover, we show how
relays' location affects the performance.\vspace{-4mm}
\end{abstract}

\section{Introduction}
Currently,  Direct-Sequence Code Division Multiple Access (DS-CDMA)
wireless networks is widely deployed \cite{80211}, such as in IEEE $802.11$b. At the link layer of such
networks, the Automatic Repeat Request (ARQ) protocol is usually
used to ensure the reliable delivery of packets with Cyclic
Redundancy Check (CRC) to check whether the received packet has
errors. If the error in the received packet is detected by CRC, the
erroneous packet is discarded and  retransmission is requested by
the receiver. There are many possible causes for errors, such as a
non-ideal state of channel between the transmitter and receiver,
collision of  packets in random access networks and so on. To
address the packet collision problem, IEEE $802.11$ wireless
networks employ Carrier Sense Multiple Access(CSMA) and the
Request-To-Send(RTS)/Clear-To-Send(CTS) mechanisms at the MAC
layer \cite{80211}. However, retransmissions are not completely
avoided because erroneous receptions still happen. Therefore, ARQ is
employed to guarantee quality of service (QoS). ARQ with a limit on
the maximum number of retransmissions, called truncated ARQ, is
applied to reduce the delay and buffer size \cite{truncatedARQ}. In
truncated ARQ, if a packet still has errors after being
retransmitted for the maximum number of times defined, the packet
will be discarded and a packet loss is announced. The truncated ARQ
can improve the packet error rate (PER), but it cannot achieve
throughput gains. This is because more transmission time is required
for better PER performance. Moreover, if the channel between source
and destination is in a poor state, the performance of throughput
cannot be improved with the increases in number of retransmissions. 

Recently, a partial packet recovery scheme was proposed to improve
the throughput performance in \cite{PPR}. In the traditional ARQ
scheme, the entire packet is  retransmitted even though there may be
only one error in the packet. The basic idea behind partial packet
recovery is to retransmit only the erroneous bits if a received
packet cannot pass the CRC. In the partial packet recovery scheme,
the receiver knows probably error bits' position in the received
packet with a unreliable decoded bits detection. The threshold
method can be applied in the unreliable decoded bits detection.
According to the results of unreliable decoded bits detection, the
receiver feeds back a request message to the transmitter so that the
transmitter is aware of which part of the packets needs
retransmission. The size of request message feedback from the
receiver to the transmitter is a key issue for the partial packet
recovery. When the request message costs too many bits to feed back
from the receiver, the throughput will decrease. A feedback strategy
based on the cost evaluation is used to solve the feedback request
issue.

Cooperative transmission techniques can provide diversity gains
through relays in the fading wireless
channel \cite{Laneman,Erkip,KJRLiu}.
In \cite{beyondbits}, a
cooperative packet recovery scheme is proposed. It requires
retransmission of the entire packet, and combines confidence
information across multiple copies of a packet from multiple access
points which are connected by wired Ethernet. In fact, this is
equivalent to a multiple antenna receiver scheme without the
assistance of relay. In \cite{Lin}, the truncated cooperative ARQ
scheme is proposed to obtain throughput gains by the relay-assisted
method, where sources and relays use an orthogonal space-time block
code (STBC) to retransmit the entire packets. However, this scheme
requires close synchronization of the source and relays for STBC to
work, and coordinating different transmitters in the wireless
network can be difficult.

To overcome the synchronization challenge, Interleave-division
multiple-access (IDMA) has the advantage that it works in an
asynchronous cooperative communication network scenario \cite{Fang}.
IDMA provides a good interference cancellation performance.
Moreover, the Multi-User Detection (MUD) in IDMA has a linear
complexity, implying a lower cost than the MMSE-based MUD which has
polynomial complexity in CDMA \cite{Lping,Liu,Leung}. In
\cite{Jeff}, a scenario is described where multiple
source-destination pairs are assisted by multiple common relays
based on IDMA. {The work in \cite{Jeff} shows} that IDMA
relays at different locations provide different diversity gains for
the multiple source-destination pairs.

In this paper, we propose a relay-assisted partial packet recovery
scheme. {In our scheme, IDMA is applied as a partial packet recovery
method. We not only bring diversity gains into the partial packet
recovery with relays, but also take advantage of IDMA for recovering
multiple erroneous packets. The proposed IDMA partial packet
recovery can be used to recover the erroneous packet under the
following scenario: a wireless network that is under heavy load may
have to handle more than one corrupted packet at the same time slot.
An example of this scenario is when CSMA and RTS/CTS fail to avoid
the collision between two source packets. If both source packets are
intending for the same destination, the receiver at the destination
will be required to handle the partial packet recovery for more than
one packet at the same time. Hence, more than one packets need to
recover at the destination in the partial packet recovery scheme. To
recover the multiple erroneous packets, the IDMA MUD may
concurrently receive and separate all retransmitted partial packets
from multiple relays for multiple source's packet. In addition, the
asynchronous property of iterative chip-by-chip MUD mechanism in our
proposed IDMA scheme enables the receiver to extract multiple
partial packets of different sizes. Because the size of different
retransmitted partial packets may be different in the case of
multiple partial packets recovery. The simulation results show that
the proposed scheme outperform the traditional ARQ. }

This paper is organized as follows: In Section II, we introduce the
system model. In Section III, we present the proposed IDMA-based
partial packet recovery scheme. Section IV, we show the simulation
result. In Section V, we provide a conclusion to this paper.

\begin{figure}
 \center
  \includegraphics[width=0.3\textwidth]{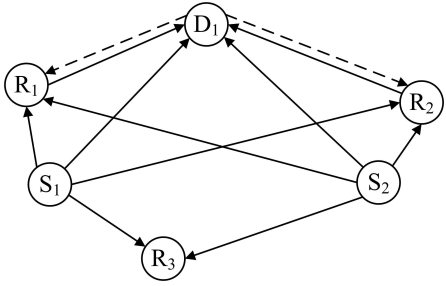}\vspace{-3mm}\\
  \caption{Relay-assisted partial packet recovery. The
solid lines denote the data transmission between nodes, the dashed
lines the feedback request from destination for retransmission. To
assist $S_{1}$ and $S_{2}$, $R_{1}$ and $R_{2}$ are the selected
best relays to  respond to the feedback
request.}\label{partialsystemmodel}\vspace{-5mm}
\end{figure}

\vspace{-4mm}

\section{System Model}
Assume that we have $K$ sources, one destination and $U$ relay nodes
in the wireless communication network. Figure
\ref{partialsystemmodel} shows the scenario with $K=2$ sources, a
destination  and $U=3$ relays with different locations. $S_{1}$ and
$S_{2}$ are the sources, and $D_{1}$ is the destination. $R_{1}$,
$R_{2}$ and $R_{3}$ are a group of relay candidates to assist in
partial packet recovery. The roles of all nodes are fixed in the
network, i.e., relay nodes will not act as a source node at
different time slot. Each node works in the half-duplex mode and
BPSK is used for the modulation.   The relay-assisted partial packet
recovery protocol divides the operation into three time slots. In
the first time slot, source nodes transmit their signal to a group
of relays and the destination. All transmissions in the first time
slot are based on CDMA. This assumes that CSMA and RTS/CTS have not
prevented more than one active source to transmit at the same time
in the network.
In the second time slot, the destination uses CRC to check whether
the received packets have errors. If no error is found, the
destination broadcast an ACK message to the sources and the group of
relays. In the third time slot, after seeing the ACK message,
sources will proceed to transmit the next packet in their queues,
and relays will erase the  packet received during the first time
slot and proceed to receive the next packet. On the other hand, if
CRC error is detected at destination in the second time slot, the
destination broadcasts a feedback request message on the feedback
channel. This request message includes the information describing
which part of the packet needs retransmission. The CRC at relays and
destination is assumed to be perfect error detection. Assume that
the feed back channel is error-free. The best relay for each
source is assumed to be known in this paper, and the relay has no
error in decoding the source's packet, as achieved by the CRC at its
receiver.Hence, $K$ relays assist $K$ sources to recovering
erroneous packets in our system model. Note that the relay is
equipped with IDMA's transmitter.  In the third time slot, the relay
responses the request, the received signal at the destination is
given by:
\begin{equation}
Y^{III}_{D}=\sum_{u=1}^{K}\sqrt{H_{RuD}P_{Ru}}X_{Ru}+N_{D}^{III},\label{dIII}
\end{equation}
where {let $I(u)$ denote the length of partial packet
transmitted by $u$-th relay}, with
 $Y^{III}_{D}=\{y^{III}_{D}(j),
j=1,2,\ldots,max(I(u))\}$. $X_{Ru}=\{x_{Ru}(j-d_{u}),
j=1,2,\ldots,I(u)\}$ denotes the unit power signal generated by the
IDMA transmitter at relay $u$, $\{d_{u}, u=1,2,\ldots,K\}$ denote
the delay variables for different partial packets.  $\sqrt{H_{RuD}}$
is the channel gain from relay $u$ to destination. $P_{Ru}$ is the
transmit power at relay $u$. $N_{D}^{III}$ denotes the noise level
at destination. $N_{D}^{III}$ follows a Gaussian distribution with
variance $\sigma^{2}$.

\begin{figure}
 \center
  \includegraphics[width=0.5\textwidth]{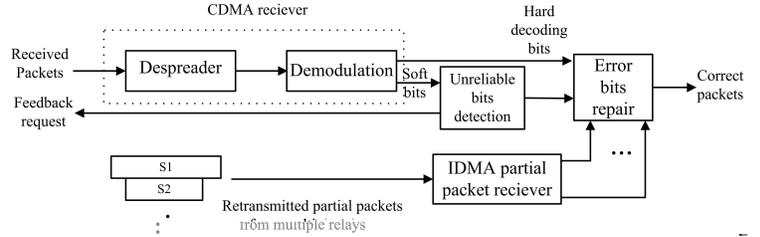}\vspace{-3mm}\\
  \caption{CDMA receiver with IDMA partial recovery module}\label{partialIDMAreceiver}\vspace{-5mm}
\end{figure}

{Figure \ref{partialIDMAreceiver} shows the structure of the CDMA
receiver at the destination which is equipped with an IDMA partial
recovery module. In Figure \ref{partialIDMAreceiver}, the hard
decoding bits and the soft bits are both outputs from the
demodulation unit. The soft bits can provide information about the
confidence level of hard decoding. The unreliable bits detection
block in Figure \ref{partialIDMAreceiver} uses the confidence
information to detect unreliable bits in the received packet. It
then feeds back a retransmission request for these unreliable bits.
Multiple partial packets retransmitted by multiple relays are
received by the IDMA partial packet receiver, Which utilizes the
chip-by-chip multiuser detection to separate the different sizes of
partial packets, denoted by $S1$ and $S2$ blocks. The received
partial packets will be input to the error bits repair block, where
the unreliable bits will be replaced by the retransmitted bits and
the multiple partial packets recovery is finally completed. }

\section{Partial packet recovery with IDMA method}
The partial packet recovery with IDMA method is activated only if
the received packet is detected to have the error in CRC. The
improvement of throughput brought by partial packet recovery is due
to save the time to retransmit the correct bits. Hence, we need to
know which part of bits are unreliable (``bad"), so the unreliable
bit detection unit is used for this purpose. Here, the concept of
soft bits and confidence values is applied to detect error bits.
The confidence values measure how much reliable of
decoding. The error bits detection is based on threshold method. If
the confidence value of a bit is larger than the preset value of
threshold, the destination recognizes a decoded bit as reliable.
Otherwise, the destination starts to determine a feedback request
strategy. The feedback request strategy is designed for
minimizing the cost of retransmission, and consequently decides a
list of retransmitted bit's indices in a packet.

 Then, this list is broadcasted on the feedback
channel. According to this list, the multiple best relays for
multiple source's packets transmit the multiple partial packet with
IDMA. In the case of multiple packets recovery, the lists
correspond to multiple packets feedback
 sequentially.
 After receiving these multiple partial packets, the destination
repairs the ``bad" bits in multiple packet to recover correct
packets.
\subsection{Soft bits and confidence values}
\begin{figure}
 \center
  \includegraphics[width=0.2\textwidth]{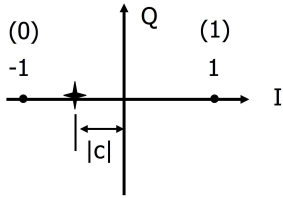}\vspace{-3mm}\\
  \caption{Soft bit in the BPSK constellation. circle points on I-axis denote constellation points, star point the received soft bit $c$. $|c|$ is the absolute value of c, which indicates the confidence value}\label{softbitconstellation}\vspace{-5mm}
\end{figure}

The concept of soft bits in partial packet recovery is similar to
the soft decoding \cite{PPR,beyondbits}. A soft bit is defined
as a real number in [-1,1] corresponding to a binary bit. The
absolute value of the soft bit indicates the confidence of decoding,
as shown in Figure \ref{softbitconstellation}. In the soft decoding
case, soft bits from demodulation are forwarded to the soft decoding
unit, and these soft bits will be discarded after the decoding. In
partial packet recovery, the soft bits are forwarded up to the link
layer for ARQ . The confidence value is a metric which measures the
reliability in the correctness of the decoded bit. Assume that the
received signal $\{y\}$ is modeled by:
\begin{equation}
y(j)=hx(j)+n(j), \ \ j=1,2,...,L
\end{equation}
where $x(j)$ is the CDMA transmitted signal, $n(j)$ denotes the
thermal noise, $h$ is the channel coefficient. Let the transmitted
BPSK symbol represented by $d(i)\in \{-1, +1\}, i=1,2,...,W$, $W$ is
the data length. ${d(i)}$ is spread by a spreading sequence
$\textbf{v}$ with the length of $V$. The spreading process is given
as: $d(i)\textbf{v}\rightarrow x(j), L=W\times V $.  Let $c(i)$
denote the output from demodulation without hard decisions. For
simplicity, we take the first BPSK symbol as an example to
illustrate the concept. After the despreading and demodulation in
Figure \ref{partialIDMAreceiver}, the first soft bit is given by
$c(1)=\frac{\sum_{j=1}^{V}v(j)y(j)}{V}$, where the numerator is the
summation over all chips related to the first BPSK symbol, the
denominator $V$ is for normalization. Figure
\ref{softbitconstellation} shows the soft bit $c$, denoted by the
star point, in the constellation. In fact, the Euclidean distance
between the $c$ and the origin is the amount of confidence values.
For example, the confidence value of the first bit can be obtained
by: $|c(1)|=|\frac{\sum_{j=1}^{V}v(j)y(j)}{V}|$.

\subsection{Unreliable decoded bits detection}\label{unreliable_detection}

\begin{figure}
 \center
  \includegraphics[width=0.35\textwidth]{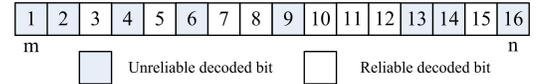}\vspace{-3mm}\\
  \caption{Unreliable decoded bits in a $16$-bit packet}\label{thresholdpacket}\vspace{-5mm}
\end{figure}

To implement the idea of partial packet recovery,  probably error
bits in a received packet should be estimated at destination so that
this part of error bits can be requested to retransmit. The
destination can utilize the confidence value provided by physical
layer to make decisions on which bits is probably error and need to
retransmit. If the absolute value of soft bits is smaller, the
confidence about
 decoding for correctness is less. That is, the high probability of error
decoding is with the bits which having smaller absolute value of
soft bits. The threshold method in \cite{PPR} is adopted to detect
unreliable decoded bits. Instead of Hamming distance in \cite{PPR},
we measure the soft bit and confidence value with Euclidian
distance. Let $T$ denotes a preset threshold of confidence value. If
a bit with the confidence value $|c|$, $|c|>T$, this bit is labeled
as a good bit. Otherwise, this bit is labeled as a bad (unreliable)
bit, retransmission of this bit is requested.  As an example, a
$16$-bit packet with unreliable decoded bits detection is
illustrated in Figure \ref{thresholdpacket}. The confidence value of
each bit in the packet is obtained by the soft bit, which is
outputted from the demodulation in the physical layer. The
unreliable decoded bits detection can be implemented in the link
layer, and the information of confidence is conveyed from physical
layer to link layer. Unreliable decoded bits are detected by
comparing confidence values with the threshold. In Figure
\ref{thresholdpacket}, the indexes of decoded bits with confidence
value lower than the threshold are $1, 2, 4, 6, 9, 13, 14$ and $16$.
Only these unreliable bits are requested to retransmit.

\subsection{Recursive algorithm of feedback request strategy}

\begin{figure}
 \center
  \includegraphics[width=1.5in,height=1.5in]{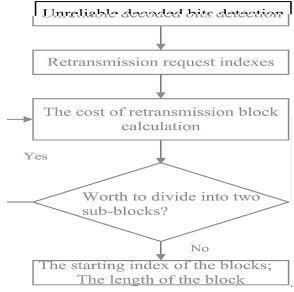}\vspace{-3mm}\\%
  \caption{Flow chart of recursive algorithm}\label{flowchart_errordetection}\vspace{-5mm}
\end{figure}

Beyond simple ACK/NACK ARQ, partial packet recovery requires the
feedback of the indexes of the unreliable bits in a packet. If this
information is large, the cost of the feedback request is large and
the throughput performance is degraded. Hence, the feedback request
strategy needs to be designed carefully. We adopt the cost-based
method, which is proposed in \cite{PPR}, to design a recursive
algorithm. The flow chart of our algorithm is shown in Figure
\ref{flowchart_errordetection}. First, the unreliable bits in a
decoded packet are detected and indexes of the unreliable bits are
obtained. Let the set $A$ denote the index set of the unreliable
bits. If the information in the index set is larger than the
retransmission of an entire block which contains both reliable bits
and unreliable bits, the retransmission of entire block is
preferred. Let $m$ and $n$ represent the starting index and the
ending index in a retransmit block in Figure \ref{thresholdpacket},
respectively. {$m, n\in A.$} Assume a packet has $L$ data bits. The
cost of retransmitting the entire block which contains all bits from
the $m$-th position to the $n$-th position in a packet is given by:
\begin{equation}
C_{I}=2\log_{2}L+n-m+1,\label{CI}
\end{equation}
where the starting index of the block needs $\log_{2}L$ bits to
describe, and the length of block also needs $\log_{2}L$ bits in the
feedback request. The number of retransmitted bits is $n-m+1$. So
the total cost is $2\log_{2}L+n-m+1$ bits. Similarly, the cost of
dividing the entire block into two sub-block is obtained by:
\begin{equation}
C_{II}=2\log_{2}L+n'-m+1+2\log_{2}L+n-m'+1,\label{CII}
\end{equation}
where $m'$ and $n'$ are the new starting index and new ending index
for the division from a block into two sub-blocks. We use the
following criterion to select the  $m'$ and $n'$ in a block:
{
\begin{equation}
m', n'\in A, m \leq n'\leq m' \leq n, s.t. \arg
max(m'-n'-1),\\\label{division_selection}
\end{equation}
where $(m'-n'-1)$ indicates that there are ($m'$-$n'$-1) reliable
bits between the $m'$-th and the $n'$-th unreliable bits. To compare
two options in (\ref{CI}) and (\ref{CII}) , we need to
 select $m'$ and $n'$ for minimizing the cost $C_{II}$ in (\ref{CII}). For the option in
 (\ref{CII}), $(m'-n'-1)$ reliable bits are not retransmitted, as the
entire block, which is from the $m$-th to $n$-th bits, is divided
into two sub-blocks, which include the $m$-th to $n'$-th bits, and
the $m'$-th to $n$-th bits, respectively. In
(\ref{division_selection}), the cost $C_{II}$ can be minimized by
maximizing $(m'-n'-1)$. } In our recursive algorithm, a decision
whether it is worth to divide a block into two sub-blocks is made
according to this cost. The cost is evaluated between the two
options by:$min(C_{I},C_{II})$,where we select the option with the smaller cost. Our recursive
algorithm keeps running on all resulted blocks after each recursion.
Once the cost does not change between iterations, the recursive
algorithm stops, and the final feedback request strategy is decided.
The starting index and length of each retransmission block is
broadcasted in the feedback channel.

\subsection{IDMA in partial packet recovery}

The retransmission request is broadcasted in the feedback channel.
This request can be received by the sources and relays.
To response the request, multiple relays apply IDMA method to
transmit multiple partial packets to the destination for the
recovery. If no relay responses the request, the source perform the
entire packet retransmission. In Figure \ref{partialsystemmodel},
the IDMA partial packet receiver applies asynchronous iterative
chip-by-chip MUD to decode the multiple partial packets from
multiple relays. The Log Likelihood Ratio (LLR) output from
asynchronous iterative chip-by-chip MUD is given as follows:
\begin{equation}
LLR(x_{Ru}(j))=\frac{2\sqrt{H_{RuD}P_{Ru}}(y^{III}_{D}(j)-E(\eta_{Ru}(j)))}{Var(\eta_{Ru}(j))},\label{LLR}
\end{equation}
where
\begin{equation}
E(\eta_{Ru}(j))=E(y^{III}_{D}(j))-\sqrt{H_{RuD}P_{Ru}}E(x_{Ru}(j-d_{u})),\label{ERu}
\end{equation}
\begin{equation}
Var(\eta_{Ru}(j))=Var(y^{III}_{D}(j))-H_{RuD}P_{Ru}Var(x_{Ru}(j-d_{u})),\label{VarRu}
\end{equation}
\begin{equation}
E(y^{III}_{D}(j))=\sum_{u=1}^{K}\sqrt{H_{RuD}P_{Ru}}E(x_{Ru}(j-d_{u})),\label{Ey}
\end{equation}
and
\begin{equation}
Var(y^{III}_{D}(j))=\sum_{u=1}^{K}H_{RuD}P_{Ru}Var(x_{Ru}(j-d_{u}))+\sigma^{2}.\label{Vary}
\end{equation}
{(\ref{LLR}) detects the received signal from the $u$-th relay in
multiple packets signal $\{y^{III}_{D}\}$. (\ref{ERu}) and
(\ref{VarRu}) are respectively the mean and variance of the
interference for the received signal from the $u$-th relay.
(\ref{Ey}) and (\ref{Vary}) are the mean and variance of the
received signal from $u$-th relay.}
 IDMA MUD estimates the statistic of the interference through the iterations.
The update rule for the estimation in each iteration for the $u$-th
partial packet is given by:
\begin{eqnarray}
E(x_{Ru}(j-d_{u})) =\nonumber \\
\left\{
\begin{array}{ll}
\tanh(\frac{I_{MUD}(x_{Ru}(j-d_{u})}{2}), & \mbox{if}\ 1\leq j-d_{u}\leq N_{u},\\
0, & \mbox{otherwise};
\end{array}
\right.
\end{eqnarray}
\begin{eqnarray}
Var(x_{Ru}(j-d_{u})) = \nonumber \\
\left\{
\begin{array}{ll}
1-E^{2}(x_{Ru}(j-d_{u})),&  \mbox{if}\ 1\leq j-d_{u}\leq N_{u},\\
0, & \mbox{otherwise},
\end{array}
\right.
\end{eqnarray}
where $I_{MUD}(x_{Ru}(j-d_{u}))$ denotes the a priori information
provided by the despreader for MUD in IDMA receiver. After decoding
the partial packet with IDMA, the decoded bits in the partial packet
is used to replace the unreliable bits in the error bits repair
block in Figure \ref{partialIDMAreceiver}. Finally, repaired bits
and reliable bits constitute a recovered packet. If the recovered
packet is still in error, the partial packet recovery with IDMA
method is activated again. Similar to truncated ARQ, the request for
partial retransmission may keep running until the recovered packet
passes the CRC check or until the maximum number of retransmission
is reached.

\vspace{-3mm}

\section{Simulation}

\vspace{-2mm}

\begin{table}
  \centering
\caption{Selected confidence value thresholds in the simulations
}\label{threshold}
  \begin{tabular}{|l|l|l|l|l|l|l|l|}
    \hline
    SNR (dB)    &-5  &0  &5 &10 &15 &20 &25\\
    \hline
    $T (\times10^{-6})$ & $0.9$ & $1$ & $1.5$ & $2.5$& $4$& $6$& $11$\\
    \hline
  \end{tabular}\vspace{-5mm}
\end{table}

\begin{figure}
 \center
  \includegraphics[width=0.45\textwidth,height=0.3\textwidth]{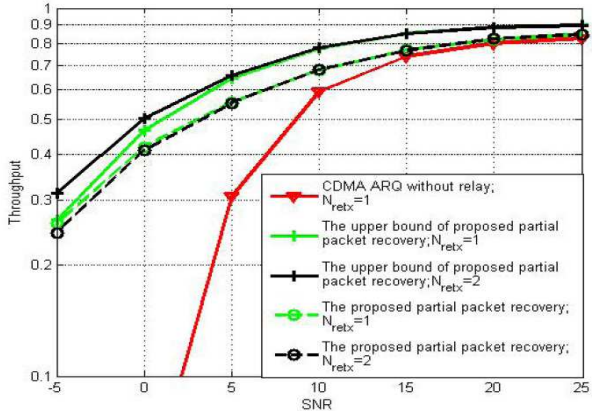}\vspace{-5mm}\\
  \caption{Throughput performance of IDMA partial packet recovery. The distance between sources to the destination is 100 meters. The distance between relays to destination is 50 meters.}\label{throughputIDMA50}\vspace{-7mm}
\end{figure}

We simulate a wireless network, which consists of two sources, one
destination, and three relay candidates over flat Rayleigh fading
channels. The length of data packet is $128$ bits. The number of
iterations at the IDMA MUD receiver are $10$. {The path loss
exponent is set to be $4$. The distance between two sources and a
destination is fixed at $100$ meters.} Let $B_{correct}$ denote the
total number of correctly received bits in the whole transmission,
$B_{T}$ the total number of transmit bits by sources and relays ,
and $B_{feedback}$ the total number of bits for feedback request in
partial packet recovery. For the performance's comparison, we define
the throughput with ACK/NACK ARQ scheme
as:$\frac{B_{correct}}{B_{T}}$.
Here, the bits used for ACK/NACK are ignored.
 The throughput with
relay-assisted IDMA partial retransmission is defined
as:$\frac{B_{correct}}{B_{T}+B_{feedback}}$, where $B_{feedback}$
includes all information describing which part of bits requested to
retransmit. This is decided by feedback request strategy. There are
two types of curves provided in our simulation: one is the upper
bound of throughput for the proposed scheme; the other is the
throughput, where the threshold method in Section
\ref{unreliable_detection} is adopted for unreliable decoded bits
detection. Meanwhile, we compare the performance of throughput with
the proposed scheme under different maximum times of retransmission
and different location settings. Let $N_{retx}$ denote the maximum
number of retransmission, and $d_{RD}$ the distance between relays
and destination. The upper bound curves assume that the unreliable
decoded bits are perfectly detected so that the destination has a
perfect knowledge of erroneous bits' position in the packet. In the
simulation, for the unreliable bit detection threshold, we select
the thresholds of confidence values listed in table \ref{threshold}
under different SNR conditions in the simulation. All values of
threshold in table \ref{threshold} are obtained by accumulative
simulations. In addition, the two best relays are selected to assist
the two sources, respectively. Figure \ref{throughputIDMA50} shows
the improvement in the throughput with the IDMA relay-assisted
partial packet recovery, over the traditional ARQ with entire packet
retransmission in CDMA networks.

\begin{figure}
 \center
  \includegraphics[width=0.45\textwidth,height=0.3\textwidth]{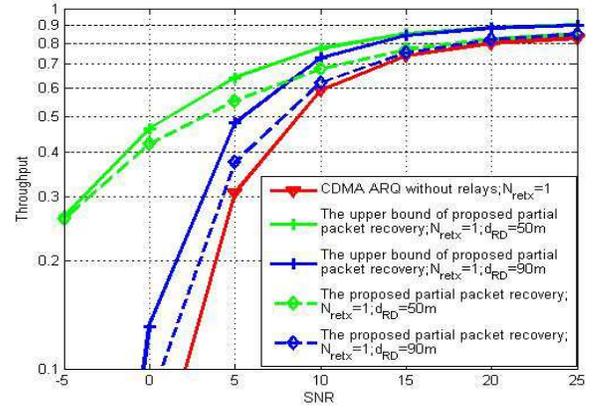}\vspace{-5mm}\\
  \caption{Throughput performance for different relay locations. Two cases of location setting: 1. Distance between relays and destination of 50 meters; 2. Distance between relays and destination of 90 meters.}\label{IDMAdifferentlocation}\vspace{-7mm}
\end{figure}

In Figure \ref{throughputIDMA50}, the two best relays are located in
the middle of sources and destination. It has been shown that IDMA
relay-assisted partial packet recovery has an advantage over the
traditional ARQ entire packet retransmission scheme in CDMA network.
Figure \ref{IDMAdifferentlocation} illustrates that the location of
the best selected relay can affect the performance of IDMA
relay-assisted partial packet recovery. Two cases of location
setting are simulated for comparison: distance between relays and
destination of $50$m, and of $90$m. When the relays are far away
from the destination, the performance is degraded because the
channels between relays and the destination degrade, and there is a
performance loss compared to the performance of the upper bound
because there can be errors in the detection of the unreliable bits
using the confidence threshold.
\vspace{-3mm}

\section{Conclusion }\vspace{-2mm}

A relay-assisted partial packet recovery scheme using IDMA is
proposed in this paper. We use relays to provide diversity gain for
retransmitting the partial packet. The transmission of
relay-assisted partial packet recovery is activated only when errors
are detected in the received packet at the destination. In our
scheme, multiple relays at different locations can assist to recover
multiple source packets using IDMA. The asynchronous IDMA iterative
chip-by-chip multiuser detection is adopted to decode multiple
partial packets. Further, we adopt the concept of confidence value
to detect the unreliable bits in a packet, and the threshold
detection method is introduced. To minimize the feedback request
overhead, we design a recursive algorithm based on cost evaluation
to determine the retransmission strategy. Simulation results show
the throughput upper bound of the proposed scheme, and the
performance improvement of the proposed scheme over the traditional
ARQ in wireless CDMA networks. In the optimal case, the performance
upper bound shows that there is at least a $7$dB gain with our
proposed scheme for a given throughput value. For the detection of
unreliable bits, the proposed scheme can obtain at least a $4$dB
gain under low SNR conditions, without optimization of the threshold
setting. To achieve more accurate detection of unreliable bits, more
sophisticated detection algorithm will be developed in future
work.\vspace{-3mm}

\bibliographystyle{IEEEbib}

\begin{thebibliography}{1}\vspace{-1mm}
\bibitem{80211}ISO/IEC, ``Wireless LAN Medium Access Control (MAC) and Physical Layer (PHY) Specifications," ANSI/IEEE Std 802.11, 1999.
\bibitem{truncatedARQ} E. Malkamaki and H. Leib,``Performance of
truncated type-II hybrid ARQ schemes with noisy feedback over block
fading channels,"{\em IEEE Transactions On Communications}, vol. 48,
no. 9, pp. 1477 - 1487, Sept. 2000.
\bibitem{PPR}K. Jamieson and H. Balakrishnan, ``PPR: Partial packet recovery for wireless
networks," in {\em ACM SIGCOMM}, pp. 409-420, Kyoto, Japan,
Aug.2007.
\bibitem{Laneman} J. N. Laneman, D. N. C. Tse and G. W. Wornell,``Cooperative
diversity in wireless networks: Efficient protocols and outage
behavior ", {\em IEEE Transactions On Information Theory},  vol. 50,
no. 12, pp. 3062 - 3080, Dec. 2003.


\bibitem{Erkip} A. Sendonaris, E. Erkip, and B. Aazhang, ``User cooperation diversity  part I:
System description", {\em IEEE Transactions On Communications}, vol.
51, no. 11, pp. 1927 - 1938, Nov. 2003.

\bibitem{KJRLiu} K. J. R. Liu, A. K. Sadek , W. Su, and A. Kwasinski, {\em Cooperative Communications and
Networking}, Cambridge University Press, Cambridge, UK, 2008.


\bibitem{beyondbits}G. R. Woo, P. Kheradpour, D. Shen, and D. Katabi, ``Beyond the bits: cooperative packet recovery using physical layer information," in {\em ACM MobiCom'07},pp.147-158, Montr\'{e}al, Qu\'{e}bec,
Canada, Sept. 2007.
\bibitem{Lin} L. Dai and K. B. Letaief, ``Throughput maximization of Ad-hoc wireless networks using adaptive
cooperative diversity and truncated ARQ," {\em IEEE Transactions On
Communications}, vol. 56, no. 11, pp. 1907 - 1918, Nov. 2008.
\bibitem{Fang} Z. Fang, L. Li, and Z. Wang,``An interleaver-based asynchronous cooperative diversity scheme for wireless relay networks",
{\em  IEEE International Conference on Communications}, pp.
4988-4991, Beijing, China, May, 2008.
\bibitem{Lping} L. Ping, L. Liu, K. Y. Wu, and W. K. Leung, ``Interleave division multiple access,"{\em IEEE Transactions On
Wireless Communications}, vol. 5, no. 4, pp. 938 - 947, Apr. 2006.

\bibitem{Liu} L. Liu and L. Ping, ``A comparative study on low-cost multiuser detectors", {\em  IEEE International Conference on Communications},
pp. 4947-4952, Istanbul, Turkey, Jun. 2006.

\bibitem{Leung} W. K. Leung, L. H. Liu, and L. Ping ,   ``Interleaving-based multiple access and iterative chip-by-chip multiuser detection",
{\em IEICE Transaction On Communications},  E86-B (12): 3634-3637,
2003.

\bibitem{Jeff}Z. Luo, D. Gurkan, Z. Han, A. K. Wong, and S. Qiu, ``Cooperative Communication based on IDMA",
in Proceedings of {\em the 5th International Conference on Wireless
Communications, Networking and Mobile Computing}, Beijing, China,
Sept. 2009.





\end{thebibliography}

\end{document}